\documentclass[12pt]{article}
\usepackage{scicite}
\usepackage{times}
\usepackage{amssymb}
\usepackage{amsthm}
\usepackage{graphicx}
\usepackage{dcolumn}
\usepackage{bm}
\newenvironment{sciabstract}{%
\begin{quote} \bf}
{\end{quote}}
\title{Emergent Hydrodynamic Bound States Between Magnetically Powered Micropropellers} 
\author
{Fernando Martinez-Pedrero,$^{1,2}$ Eloy Navarro-Argem\'i,$^{1}$ \\ Antonio Ortiz-Ambriz,$^{1}$ Ignacio Pagonabarraga,$^{1,3}$, Pietro Tierno$^{1,3,4\ast}$\\
\\
\normalsize{$^{1}$Departament de F\'isica de la Mat\`eria Condensada, Universitat de Barcelona, Barcelona, Spain}\\
\normalsize{$^{2}$Departamento de Qu\'imica F\'isica I, Universidad Complutense de Madrid, Madrid, Spain}\\
\normalsize{$^{3}$Universitat de Barcelona Institute of Complex Systems (UBICS), Barcelona, Spain}\\
\normalsize{$^{4}$Institut de Nanoci\`encia i Nanotecnologia, IN$^2$UB, Universitat de Barcelona, Barcelona, Spain}\\
\\
\normalsize{$^\ast$To whom correspondence should be addressed; E-mail:  ptierno@ub.edu.}
}
\date{}

\begin{document} 

\baselineskip24pt

\maketitle 

\begin{sciabstract}
Hydrodynamic interactions (HIs),
namely solvent mediated long-range
interactions between dispersed particles,
play a crucial role in the assembly and dynamics
of many active systems,
from swimming bacteria to
swarms of propelling microrobots.
Here we experimentally demonstrate 
the emergence of long-living hydrodynamic bound states 
between model micro-swimmers at low
Reynolds number. 
A rotating magnetic field
forces colloidal hematite microparticles 
to translate at a constant and frequency-tunable speed
close to a bounding plane in a viscous fluid.
At high driving frequency,
HIs dominate over magnetic dipolar ones,
and close propelling particles
couple into bound states by adjusting their translational speed
in order to optimize the transport of the pair.
The physical system is described by considering the 
HIs with the boundary surface and the 
effect of gravity, providing an excellent agreement with the 
experimental data for all
the range of parameters explored.
Moreover, we show that in dense suspensions, these bound states
can be extended to one-dimensional 
arrays of particles assembled by
the sole HIs.
Our results manifest
the importance of the boundary surface
in the interaction and dynamics of
confined propelling microswimmers.
\end{sciabstract}
\section*{INTRODUCTION}
Active particles
moving in viscous fluids driven by
external fields~\cite{Dre05,Tie08,Sne09,Zha09,Bri13}
or chemical reactions~\cite{Pax04,How07}, represent a rich
and growing area of research where many
emergent non-equilibrium
phenomena
can be observed~\cite{Ara13}.
The dynamics
of an ensemble of
active particles are governed
by the interplay between propulsion, 
thermal noise and
pair-interactions coupled with the 
dispersing medium.
External fields usually induce 
alignment  and fast assembly of the particles
due to dipolar forces~\cite{Mar13a}.
Subtler is the role played by
HIs,
i.e. the solvent-mediated long-range
interactions,
that can be excited by the random or directed motion
of the particles
in the fluid medium.
These interactions depend on the particle speed and angular velocities, since
the solvent couples to the particles through forces and torques. 
Thus HIs cannot be described by a simple potential term, 
in contrast to dipolar forces.

Colloidal dispersions constitute a
suitable model system to study the complex effects of
HIs at low Reynolds number, 
given the ability to
manipulate~\cite{Reic04,Mar06,Lut06,Zie09}, align~\cite{Bru12,Nag14,Wil14} or
rotate~\cite{Dil12,Kou13,Kot13} the particles with 
external fields.
More generally,
HIs are not only limited to
colloids,
but affect the dynamics of many
complex and biological systems~\cite{Eric09,Mar13}.
Examples span from the
spontaneous formation of cell vortex arrays~\cite{Rie05},
to synchronized cilia beating~\cite{Vil06,Gui07},
and the attraction and dancing of a pair of algae Volvox~\cite{Dre09}.
Thus, understanding the effect of HIs
in interacting microscale objects
is crucial in both
applied and fundamental research.

When the motion of the propelled particles 
is affected by a nearby wall, 
the confinement may have a strong influence on the system dynamics.
For example, the proximity of a surface explains the
circular path of motile {\it E. coli}~\cite{Eri05},
the attraction of model swimmers toward the surface~\cite{Ber08}, or
can be even used to steer
the particles along defined paths~\cite{Wan08,Dil11,Pot13,Led13}.
Most of the experiments based on
trapped or driven colloidal particles 
used as model system, 
have been mainly focused on the direct 
interactions between the couple,
without exploring the effect of a confining plane 
on the collective dynamics.

Here
we investigate the interactions between
a pair of model microswimmers,
composed of monodisperse
ferromagnetic hematite microparticles
driven above a plane
by an external rotating magnetic
field.
At high driving frequencies,
the spinning particles generate large hydrodynamic flow fields,
and the corresponding HIs
dominate
over the magnetic ones,
giving rise to different
dynamic states.
We report the direct experimental observation of
hydrodynamic bound states
with a long lifetime,
where two close propelling particles
adjust their
velocities in order to
couple their trajectory.
Depending on the relative position of the particles,
the bound colloids could either translate
tip to tip, or speed up
by aligning
such that
the relative position is perpendicular to the particles' long axes.
The experiments are complemented
with a theoretical model
that takes into account the presence of the bounding plane,
allowing to capture the essential physics
of the process.

\section*{RESULTS}
In our experiments, we
we use "peanut" shaped hematite
microparticles,
characterized by a long (short) axis equal to $a=2.5 {\rm \mu m}$ ($b=1.4 {\rm \mu m}$),
Figs.1(a,b).
After dilution in highly deionized water,
the particles sediment due to
density mismatch above a glass substrate,
where they display small Brownian motion
with negligible out-of plane fluctuations.
The particles have a small permanent moment
$\bm{m}$ perpendicular to their long axis.
We determine the amplitude of the
particle moment by following the
orientational angle $\theta$
of individual particles
when subjected to a static magnetic field $\bm{H}$.
Fig.1(c) shows the evolution with time of
$\theta$ for $5$ different particles and the corresponding 
fits to the experimental data,
see the Method section. 
From these fits, we find a distribution of moments $P(m)$
centered at $\langle m \rangle = 9 \cdot 10^{-16} {\rm A \, m^2}$ ,
and having a Gaussian-like shape with variance 
$\sigma_m=3 \cdot 10^{-16} {\rm A \, m^2}$.

We spin the hematite particles
by applying
a rotating magnetic field
circularly polarized in the $(\hat{x},\hat{z})$ plane,
${\bm H}(t) = H_0 (\cos{(\omega t)}{\bm e}_x- \sin{(\omega t)}{\bm e}_z)$
with angular frequency $\omega$
and amplitude $H_0$, 
Fig.1(a).
The applied modulation induces a magnetic torque
${\bm T}_m=\mu_0 {\bm m} \times {\bm H} $,
which sets the particles in rotation close to the substrate 
at an average
angular velocity ${\langle \Omega \rangle}$, with $\mu_0 = 4 \pi \cdot 10^{-7}  H \, m^{-1}$.
The solid surface breaks the spatial symmetry, and the
anisotropic particles roll close to the plane due to the rotation-translation 
hydrodynamic coupling~\cite{Gol67}.
The net drift velocity is a function of the angular
speed, $\langle v_x \rangle = b f_r \langle \Omega \rangle /2$,
being $f_r$ a small correction factor resulting from the wall 
proximity. 
From the fit in Fig.1(d), and 
using the expression of $f_r$~\cite{Ferr},
we estimate the average elevation 
of the particle to be $h = 1.0 {\rm \mu m}$.
Below a critical frequency $\omega_c$,
single particles follow the field rotation
synchronously, with
$\langle \Omega \rangle = \omega$.
For $\omega > \omega_c$
the particles follow  asynchronously the driving 
magnetic field, and  $\langle  v_x \rangle$ decreases as
the $\omega$ increases. Neglecting thermal fluctuations,
the average
rotational speed follows
$\langle \Omega \rangle=\omega (1-\sqrt{(1-(\omega_c/\omega)^2)})$~\cite{Adl46},
as shown by the blue fit in Fig.1(d).
We note that the hematite particles  display a 
weak magnetic-moment polydispersity, 
as also measured in Fig.1(c). 
As a consequence,
close to $\omega_c$,
the propellers show
a speed distribution,
$P(\langle v_x \rangle)$,
and similar particles
may display different average speeds
being in the synchronous
or asynchronous regime.

When we increase the density of the 
colloidal system, close particles 
start to interact due to 
magnetic and viscous forces. 
The driving field aligns the permanent moments of the micropropellers,
ensuring that  the particles always roll perpendicular to their long axis.
Thus, 
only the angle $\vartheta$ between $\hat{x}$-axis and the line connecting the center
of the two particles
is needed to describe the relative
orientation of the pair, Fig.1(e).
In order to classify the type of dynamic states
observed, we vary mainly the driving frequency 
that controls the average speed, and fix the field amplitude.
We find that the relative arrangements of 
pairs of particles strongly depend on 
their rotational motion.
At low driving
frequencies, $\omega \lesssim 125.7 {\rm rad s^{-1}}$, 
the propellers tend to arrange such that
the relative position is perpendicular to the particles' long axes,
in order to minimize the  magnetic energy of the couple,
first column of Fig.1(e).
In contrast
we find that, at high driving frequencies,
the fast spinning of the particles induces strong
HIs, and the propellers assemble forming a different
bound state, in which the particles
align along their long axis
during propulsion, second column in Fig.1(e),
even if dipolar interactions are repulsive in this configuration, 
as demonstrated in the next section.

We can identify the emergence of these bound states by measuring
the order parameter $\alpha=(v_{rel}/v_{cv})^2$,
based on the speed $v_{1,2}$ of the particles,
which quantifies the ratio between
the relative velocity
$v_{rel}=v_1-v_2$ and
center of velocity
$v_{cv}=(v_1+v_2)/2$ of the couple.
Figs.2(a-c) show
three representative cases illustrating 
the behavior
of a couple of propellers
with the low frequency situation
displayed in the first column.
At high frequencies, we find
that colloidal pairs tend to
either align at a finite angle ($\vartheta \in [\pi/6, \pi/3]$),
second column in Figs.2(a-c), or to arrange themselves parallel to each other
such that $\vartheta= \pi/2$,
third column in Figs.2(a-c). In the last two cases,
the two particles form a transient  hydrodynamic
state that can last more than $6$ seconds, and where the two particles
adjust their velocities such that
either $v_{cv}$
increases ($\vartheta \rightarrow 0$)
or decreases ($\vartheta \rightarrow \pi/2$),
and $v_{rel}$ almost vanishes.
We quantify the duration of these bound states by measuring the 
histogram of the distribution of time 
lapses during which $\alpha<0.01$. 
We find that with this choice, 
these states
have a longer duration at high frequencies, 
even if magnetic interactions
oppose their development, 
as shown by Figs.2(d,e).
The presence of disorder and thermal noise
in the experimental system may be at
the origin of the finite lifetime of these states. However,
we have found cases where the bound states
were observed along the whole sample area, providing a
strong indication that the observed trajectories do not
correspond to simple scattering events between the particles.

\section{THEORETICAL MODEL}
\textbf{Hydrodynamic interactions.}
We model the propelling couple as a pair
of rotating solid spheres above a wall.
The solid surface
can be accounted through an
hydrodynamic singularity placed at the same
distance  below the position of
the interface, namely a particle rotating
in the opposite sense, plus additional
stresslet and source doublets~\cite{Bla74}. 
We assume that the applied
field forces the particles to rotate at 
a given angular velocity, which may
differ from the driving frequency depending on the nature of each colloidal particle.
Thus, the flow
generated by a colloid
of radius $a$ and rotating at a prescribed angular
velocity $\Omega$,
can be estimated by considering a rotlet located at a distance $h$ from the wall:
\begin{eqnarray}
\frac{u_i}{a^3}=\frac{\epsilon_{ijk}\Omega_j r_k}{r^3}-\frac{\epsilon_{ijk}\Omega_jR_k}{R^3}+ \nonumber \\
2h\epsilon_{kjz}\Omega_j \left(\frac{\delta_{ik}}{R^3}-\frac{3R_iR_k}{R^5} \right)+6\epsilon_{kjz}\frac{\Omega_jR_iR_kR_z}{R^5} \,  .
\label{eq1}
\end{eqnarray}
Here $r$ is the position vector from the center of the particle and $R$ the position of its image.
From Eq.1 we derive 
analytic expressions for the speeds $\bm{v}_{1,2}$
for two spheres ${1,2}$ rotating with
angular speed $\Omega_{1,2}$ and at elevation $h_{1,2}$.
The complete set of equations resulting from Eq.(1) 
are given in the Supplementary 
Information file.

From the model we can first resolve the complex dynamics 
of the particles in a bound state. 
Typical trajectories of the center of velocities of the two particles forming a
pair are shown in Fig.3(a) and in MovieS4 in the Supplementary 
Information,
when driven by an external rotating field in the synchronous regime. When the particles
form the bound state, their relative distance is not fixed in one period, but the two propellers perform a periodic motion
around each other in the 3D plane. During this motion, the particle elevation and their relative distance periodically
vary thus, the type of bound orbit could depend on the initial angle and distance. This relative motion can be clearly
visualized from simulations, while being more difficult to be quantified from the experiments.

\textbf{Role of gravity in the bound state}
We also consider the effect of gravity on the dynamics of the pair of particles.
The gravitational force $F_g=\Delta \rho g V$
is balanced by the repulsive electrostatic interaction 
$F_{el}=\frac{A}{\lambda} e^{-z/\lambda}$
arising from the charge of the particle and the 
bounding plate. 
Here $V=(4/3)\pi a^3$ is the particle volume, 
$\Delta \rho$ is the density mismatch
between the particle and the suspending medium,
$A$ a prefactor that depends on the surface charge density and $\lambda$ the Debye screening length. 
For small displacement $z=h+\delta$, 
near the equilibrium elevation 
of the particle $h$,
one can derive an effective elastic 
force with a coupling 
constant given by,
$K=\frac{gV \Delta \rho}{\lambda}$.
We consider the effect of the vertical forces 
as a stokeslet, and add its contribution
to the Eq.1 in the Supplementary 
Information, where the 
resulting set of equation are given. 
These equations allow testing for the role of gravity 
in the bound state by varying the 
gravitational force $F_g=\Delta \rho g V$.
In particular, in Fig.3(b) we plot the mean distance between 
two particles in a bound state as a function of the period 
of the rotating magnetic field. 
The different curves refer to situations where 
we vary the ratio $F_{g/h}$ between $F_g$ and the viscous force $F_h=3\pi \eta \Omega a^2/4$,
being $\eta=10^{-3} \, \rm{Pa \cdot s}$ the solvent viscosity and 
we used $h \sim a$ as the particle elevation from the substrate. 
In all cases we find that, for small initial distance between the particles,  
stable bound states with no particle separation 
can be formed in absence of gravity ($F_{g/h}=0$)
or even for strong gravitational forces  ($F_{g/h}=1$).
Thus, we still observe the formation of these states
and find that the main effect of gravity 
is to reduce the three-dimensional
movement of the particles when they couple. 
The consequence of this constrained motion 
is that the relative positional angle between the particles 
changes, as shown in Fig.3(c). Starting from a configuration of $\vartheta=45^{o}$,
even small gravitational forces ($F_{g/h}\sim10^{-2}$) 
are able to stabilize the couple of particle in the tip to tip configuration,
illustrated in the second column of Fig.1(e).
This configuration is linearly unstable 
for the purely hydrodynamic model,
but it becomes stable in presence of gravity.

\textbf{Magnetic interactions.} 
We here show that the dipolar interactions 
between a pair of propellers are effectively attractive
(repulsive) when their relative position is 
perpendicular (parallel) to the particle long axis. 
The magnetic dipolar
interactions
between a pair of particles, $(i,j)$,
with moments $\bm{m}_{i}$,$\bm{m}_{j}$ and 
at a distance
${\bm r}_{ij}=|\bm{r}_i-\bm{r}_j|$,
is given by
$U_{d}=\frac{\mu_0}{4\pi}(\frac{{\bm m}_i {\bm m}_j}{r_{ij}^3}-\frac{3({\bm m}_i \cdot {\bm r}_{ij})({\bm m}_j\cdot {\bm r}_{ij}) }{r_{ij}^5})$.
Thus $U_{d}$
is maximally attractive (repulsive) for particles with
magnetic moments parallel (perpendicular) to ${\bm r}_{ij}$.
If we consider two dipoles confined on the
($x,y$) plane, and aligned by a magnetic field rotating in a perpendicular plane ($y,z$),
as shown in Fig.1(a),
we can calculate the average interaction energy between them 
by performing a time average~\cite{Mar13a}. 
We find that 
the effective potential
is attractive $\langle U_d \rangle =-\frac{\mu_0 m^2}{8 \pi (x+z)^3}$,
leading to chaining in the $(\hat{x},\hat{z})$ plane,
while being repulsive in the
perpendicular plane $(\hat{x},\hat{y})$, $\langle U_d \rangle =\frac{\mu_0 m^2}{4 \pi y^3}$.

Further, we can put forward a simple 
argument to explain the different arrangements
from the changes in magnitude of the competing forces acting on the pair. 
The attractive force between
two equal dipoles $m$ at distance
$r=2a$
is given by $F_{m}=3\mu_0 m^2/(64 \pi a^4)$,
while the viscous force generated by a single propeller
is $F_h=3\pi \eta \Omega a^2/4$,
being $\eta=10^{-3} {\rm Pa \cdot s}$ the solvent viscosity
and $h\sim a$.
Hence, for $m= 9\cdot 10^{-16} \, {\rm A m^2}$
and $a=b/2=0.7 {\rm \mu m}$,
we find that $F_h/F_m=16\pi^2\eta \Omega a^6/(\mu_0 m^2)= 0.02\Omega$.
Thus $F_h=0.6F_m$ for $\Omega = 31.4 {\rm rad \, s^{-1}}$, while $F_h=9.1F_m$ for $\Omega = 504.6 {\rm rad \, s^{-1}}$
Finally, we note that the magnetic interactions due to the permanent moments of the particles
are relatively small, and the tip-to-tip  (side-by-side) 
alignment 
gives an interaction potential $U_d=7.3 k_B T$ ($U_d=1.3 k_B T$),
i.e. much lower than the interactions
resulting from induced dipoles~\cite{Tie16}. 
Thus, while in our previous works~\cite{Tie16,fern15} the magnetic interactions 
were essential to maintain the propelling structures, here 
the dynamic states observed at large frequencies are 
bound purely by hydrodynamic interactions mediated through the confinement.

\section*{DISCUSSION}
We test the model
by measuring
the average translational speed $\langle v_{cv} \rangle$
of a pair of particle in a bound state.
In particular, we decompose this velocity in two components 
parallel ( $\langle v_{cv} \rangle \cdot \hat{x}$)
and perpendicular  ( $\langle v_{cv} \rangle \cdot \hat{y}$)
to the propulsion direction ($\hat{x}$-axis) 
imposed by the rotating field.
Both quantities, normalized with respect to the speed of a single
propeller $\langle v_0 \rangle$,
are plotted versus the positional angle $\vartheta$ in Figs.4(a,b),
and at different center to center distances $r$.
The image shows the comparison between the experimental data (symbols)
and the analytical results (continuous line)
which were plotted assuming that both particles
have the same elevation, $h \sim 1 {\rm \mu m}$,
as determined from the fits in Fig.1(d),
and are driven in the synchronous regime 
with $\Omega_{1,2}=\omega=502.6 \, \rm{rad s^{-1}}$.
The data and the fits
show that along the propulsion direction, 
the pair of particles 
decreases their average speed with the
angle $\vartheta$, being the configuration
where the relative position is perpendicular to the particles' long axes
the fastest one. The
behavior of the perpendicular component 
also illustrates the tendency of the pair
to have a higher transversal velocity 
at intermediate angles, 
in agreement with the experimental observation where 
pair of propellers were found to speed up
when placed at $\vartheta \sim 45^{0}$, 
see also Figs.2(b), second row.
Despite the approximations in the model, we obtain quantitative agreement
with the experimental data,
as the model well captures the physical 
mechanism behind the HIs in these bound states.

Self-organization induced by HIs
at low Reynolds number
is of attracting remarkable interest
due to its direct connection with
assembly and swarming in active
living
systems~\cite{Len03,Bar08,Leo10,Got10,Yeo15,Yus15,Guz16,Dri17}.
We demonstrate that our colloidal
propellers, when driven at high frequencies, 
can be organized into
metastable, elongated structures
maintained only by HIs.
Fig.5(a) shows a series of images
illustrating the assembly
process of a chain
composed by $5$ hydrodynamically coupled
propellers  driven by the
rotating field.
As shown in MovieS3 in the Supplementary 
Information,
to assemble this structure, and avoid that the particles travel outside
the field of view, the propellers were driven
first forward and later backward
when reaching the limit of the observation area. 
This change of direction was
obtained by switching the
$x-$component of the applied field, $H_x = - H_x$.
Reversing the field does not disentangle the 
formed bound state since gravity and disorder break the symmetry 
under time inversion at low Reynolds number.
From the particle positions, we determine the
angle $\vartheta$, and the velocities
$\langle v_{cv} \rangle$,
$\langle v_{rel} \rangle$ which now are averaged over
all pairs
of particles within the assembly. As shown in
Fig.5(b), after a transient regime,
all quantities stabilize
to a stationary value where
the relative speed between the composing particles approaches zero.
The composite chain is rather stable, propelling
as a compact rod
with a center of velocity
$\langle v_{cv} \rangle \sim 10 {\rm \mu m s^{-1}}$.
From Eq.1 it is also possible to calculate the net
flow field velocity generated by the assembly, which
is presented in the small inset in Fig.5(b)
as a streamplot graph.
The chain of propellers generates
a net flux toward the direction of motion,
perpendicular to the chain long axis
and focused towards the center of the assembly.
This flow field may be used to manipulate unbound
non magnetic objects in a fluid,
both pulling or pushing them by varying the propulsion
direction of the chain
through the sense of rotation of the actuating magnetic field.
Plans to explore these exciting possibilities are under way.

In conclusion,  we have investigated the
long range hydrodynamic interactions between two model
microswimmers
composed by magnetically propelled
anisotropic hematite particles.
When HIs dominate
over the dipolar one,
we observe the emergence of
cooperative states hydrodynamically bound,
where the particles
adjust their speed
by slowing down
or  speeding up
due to the
generated flow field.
Our findings may help understand
similar cooperative mechanisms occurring in
confined biological systems at low Reynolds number,
and be a starting point towards
the description of the
dynamics in dense driven particle suspensions.

\section{MATERIALS AND METHODS}
\textbf{Magnetic propellers and experimental set-up.}
Ellipsoidal hematite ($\gamma$-Fe$_2$O$_3$) particles
are synthesized by following the
technique developed by Sugimoto
and coworkers~\cite{Sug96}.
Specifically, we gradually add
to an iron chloride hexahydrate solution
($54.00 {\rm g} \, {\rm FeCl}_3 \, - \, 6$H$_2$O in $100 {\rm ml}$
of high deionized water)
a sodium hydroxide solution
($19.48 {\rm g}$ of NaOH in $90 {\rm ml}$ of Millipore water)
and stir both solutions at $75^{o} {\rm C}$.
After $5 {\rm min}$, we add to the stirring
mixture $10 {\rm ml}$ of an aqueous solution containing $0.29 {\rm g}$
of potassium sulfate
(K$_2$SO$_4$). The resulting brown mixture
is then stirred for another $5 {\rm min}$,
hermetically sealed, and left
aging in an oven at $100^{o} {\rm C}$ for $8$
days.\\
The ellipsoidal-like particles are then recovered by diluting the suspension with
high deionized water, letting the particles sediment and removing
the resulting yellowish-brown supernatant, a procedure
that is repeated several times.
In order to avoid sticking of the particles to the
glass substrate, the hematite ellipsoids are then functionalized
with sodium dodecyl sulfate (SDS).\\
The measurement cell is placed in the center of two
orthogonal pairs of coils arranged on the stage of an
optical microscope and aligned along the $x$ and $z$ axis.
To apply a rotating field in a plane, the two pair of coils are
connected with a wave generator (TGA1244, TTi) feeding a power
amplifier (AMP-1800, AKIYAMA or BOP 20-10M, Kepco) and two
sinusoidal currents with $90 \rm{degrees}$ phase-shift
are passed through the coils.
The particle dynamics are recorded by using a CCD camera
(Basler Scout scA640-74fc) working at $75$ frame per second. 
The camera is mounted on top of a
light microscope (Eclipse Ni, Nikon) equipped with
different magnification objectives ($100\times$, $40\times$)
and a $0.45\times$ TV lens adapter. The positions of the particles
are obtained from the analysis of .AVI videos recorded via
a commercial software (streampix, NORPIX).

\textbf{Determination of the magnetic moments.} We measure the distribution of magnetic moments
by following the orientation of
different hematite particles subjected to
an external static magnetic field $\bm{H}$, as shown in the
schematic in Fig.1(c).
In first approximation, 
we assume that the shape of our hematite 
particle
resembles that of an ellipsoid,
and thus use the friction coefficient of ellipsoid in water 
to describe the 
particle dynamics.
When reoriented by the external field,
the magnetic torque acting on the ellipsoid,
$\bm{\tau}_m=\mu_0 \bm{m} \times \bm{H}$,
is balanced by the viscous torque arising from its rotation
in the fluid,
$\bm{\tau}_v=-\xi_r \dot{\bm{\theta}}$.
Here $\mu_0= 4 \pi \cdot 10^{-7} {\rm H \, m^{-1}}$
and $\xi_r$ the rotational friction coefficient
of the ellipsoid. By solving the torque
balance equation, $\bm{\tau}_m+\bm{\tau}_v=0$,
and taking into account that the angle
between the permanent moment and the ellipsoid long axis is $\pi/2$, we arrive at
the equation:
\begin{equation}
\theta(t)=2 \tan^{-1}{ \left[ \tanh{ \left(\frac{t}{\tau_r}  \right)} \right]}  \, \, .
\label{Eq_torque}
\end{equation}
The rotational friction coefficient
for a prolate ellipsoid rotating around
its short axis can be written as
$\xi_r=8\pi \eta V_c f_r$,
where 
$V_c=(4 \pi a b^2)/3$
is the volume of the ellipsoid,
and $f_r$ is a small geometrical factor which
depends on the lengths of the
ellipsoid long and short axes.

\bibliography{scibib}
\bibliographystyle{Science}

\section*{Acknowledgments}
F.M.P. , A.O.A. and P.T. acknowledge support from the
ERC starting Grant "DynaMO" (No. 335040).
F.M.P. acknowledges support from the Ramon y Cajal program (RYC-2015-18495).
E.N.A. and I.P. acknowledges support from MINECO (Spain),
Project FIS2015-67837-P, DURSI Project 2014SGR-922,
and Generalitat de Catalunya under Program "ICREA  Acad\`emia".
P.T. acknowledges support from
from MINECO (FIS2016-78507-C2) and DURSI (2014SGR878).

\section*{SUPPLEMENTARY MATERIALS}
As supporting Information we provide one 
.pdf file with the derivation of the 
set of equations used to fit the experimental data in the main text.

\subsubsection*{Supporting Information Movies}
With the article are 4 videoclips as support of Figs.1,3 and Fig.4.
\begin{itemize}

\item {\bf MovieS1}(.WMV) This videoclip illustrates the
dynamics of a pair of propellers
driven by a rotating
magnetic field with amplitude $H_0 = 1500A m^{-1}$
and frequency $\omega=31.4 rad \, s^{-1}$ (left side),
and $\omega=502.6 {\rm rad \, s^{-1}}$ (right side).
The video corresponds to Fig.1(e) of the article.

\item {\bf MovieS2}(.WMV) The video illustrate the
increasing velocity of
a pair of propellers
driven by a rotating magnetic field
with amplitude $H_0=4400 {\rm A \, m^{-1}}$
and frequency
$\omega=502.6 {\rm rad \, s^{-1}}$.
The video corresponds to the middle row of
Fig.2(b) of the article.

\item {\bf MovieS3}(.WMV) Assembly and transport of
a chain composed by $5$ propellers
subjected to a rotating magnetic field with amplitude
$H_0=4400 {\rm A m^{-1}}$
and frequency $\omega= 502.6 {\rm rad \, s^{-1}}$.
During the video one component of the
rotating field is inverted ($H_x=-H_x$)
each time the collection of particles
reaches the edge of the observation area.
The video corresponds to Fig.5(a) of the article.

\item {\bf MovieS4}(.WMV) The video shows the 
trajectory of the center of a pair of particles 
in an hydrodynamic bound state obtained from numerical simulation of the 
equation of motion of the pair. 
The video corresponds to Fig.3(a) in this 
Supporting Information. 

\end{itemize}
\clearpage

\begin{figure*}[t]
\begin{center}
\includegraphics[width=0.9\textwidth]{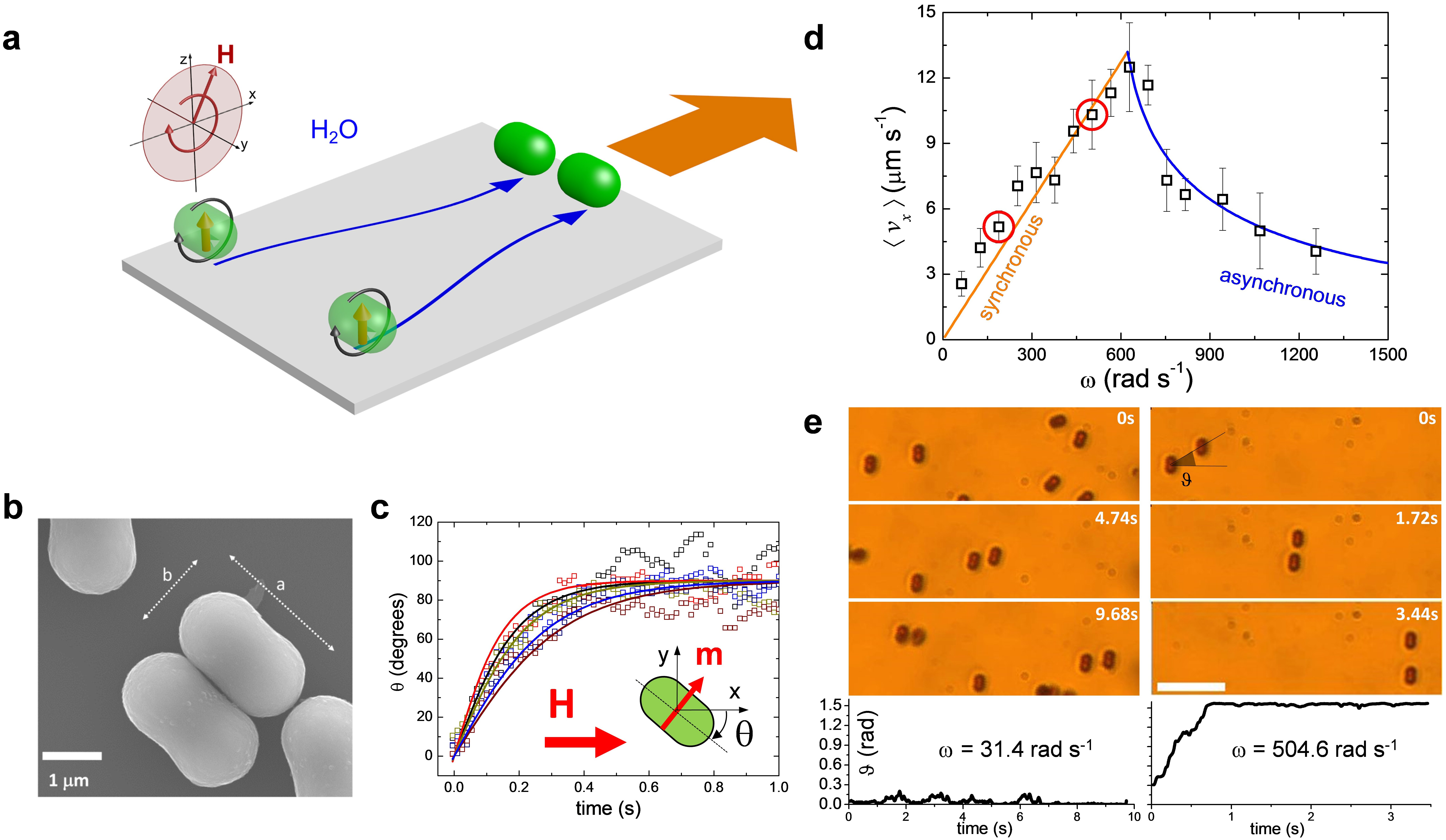}
\caption{\label{fig1} Propulsion of the hematite micropropellers. (\textbf{a}) 
Schematic showing two 
hematite particles subjected to a rotating magnetic field circularly polarized in the $(x,z)$ plane. 
(\textbf{b}) 
SEM image of the hematite particles, being
$a$ and $b$ the long and short axis,
resp. 
(\textbf{c}) 
Orientational angle $\theta$ between the particle long axis and 
an applied 
field $\bm{H}$. The particle moment 
is obtained by balancing the magnetic and viscous torque during re-orientation,
see Method section. 
(\textbf{d}) Average speed $\langle v_x \rangle$
versus driving frequency $\omega$
of one micropropeller subjected to a rotating field
with amplitude $H_0= 4400 {\rm A \, m^{-1}}$.
Orange (blue) fit denotes synchronous (asynchronous)
regime.
(\textbf{e}) Microscope images showing
a pair of propelling
particles at frequencies $\omega=31.4 {\rm rad \, s^{-1}}$ (left) and
$\omega=502.7 {\rm rad \, s^{-1}}$ (right).
Scale bar is $10 \, {\rm \mu m}$,
see MovieS1 in the Supplementary 
Information.
Bottom panel shows the evolution with time of
the relative positional angle $\theta$.}
\end{center}
\end{figure*}
 
\begin{figure*}[tbp]
\begin{center}
\includegraphics[width=\textwidth]{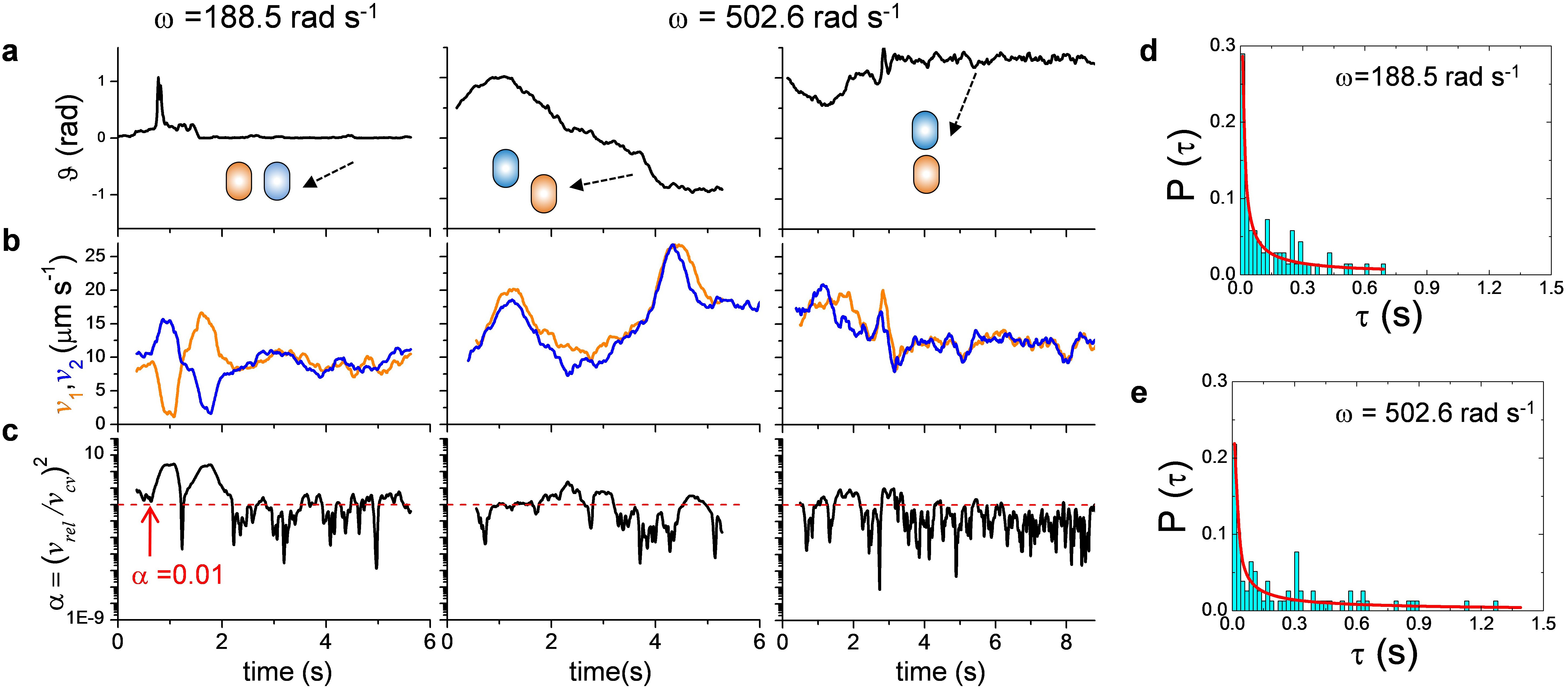}
\caption{\label{fig2} Observation of the bound states. \textbf{(a-c)} Sequence of graphs illustrating the evolution of the
angle $\vartheta$ (a),
particle velocities $v_1,v_2$ (b)
and normalized relative velocity squared
$\alpha=(v_{rel}/v_{cv})^2$ (c)
versus time
for a pair of propellers driven by a rotating field
with amplitude $H_0=4400 {\rm A \, m^{-1}}$
and at two different frequencies.
The first column refers to $\omega=125.7 {\rm rad \, s^{-1}}$,
second and third columns to $\omega=502.6 {\rm rad \, s^{-1}}$.
The increase in speed of the pair of propellers 
is shown in MovieS2 in the Supplementary 
Information.
\textbf{(d,e)} Probabilities $P(\tau)$
of times $\tau$ where the pair of propellers
have $\alpha<0.01$
for frequencies $\omega=125.7 {\rm rad \, s^{-1}}$ (d),
and $\omega=502.6 {\rm rad \, s^{-1}}$ (e).
The continuous red line are 
fits to the data of an 
algebraic function as a guide to the eye.}
\end{center}
\end{figure*}

\begin{figure*}[t]
\begin{center}
\includegraphics[width=0.5\textwidth]{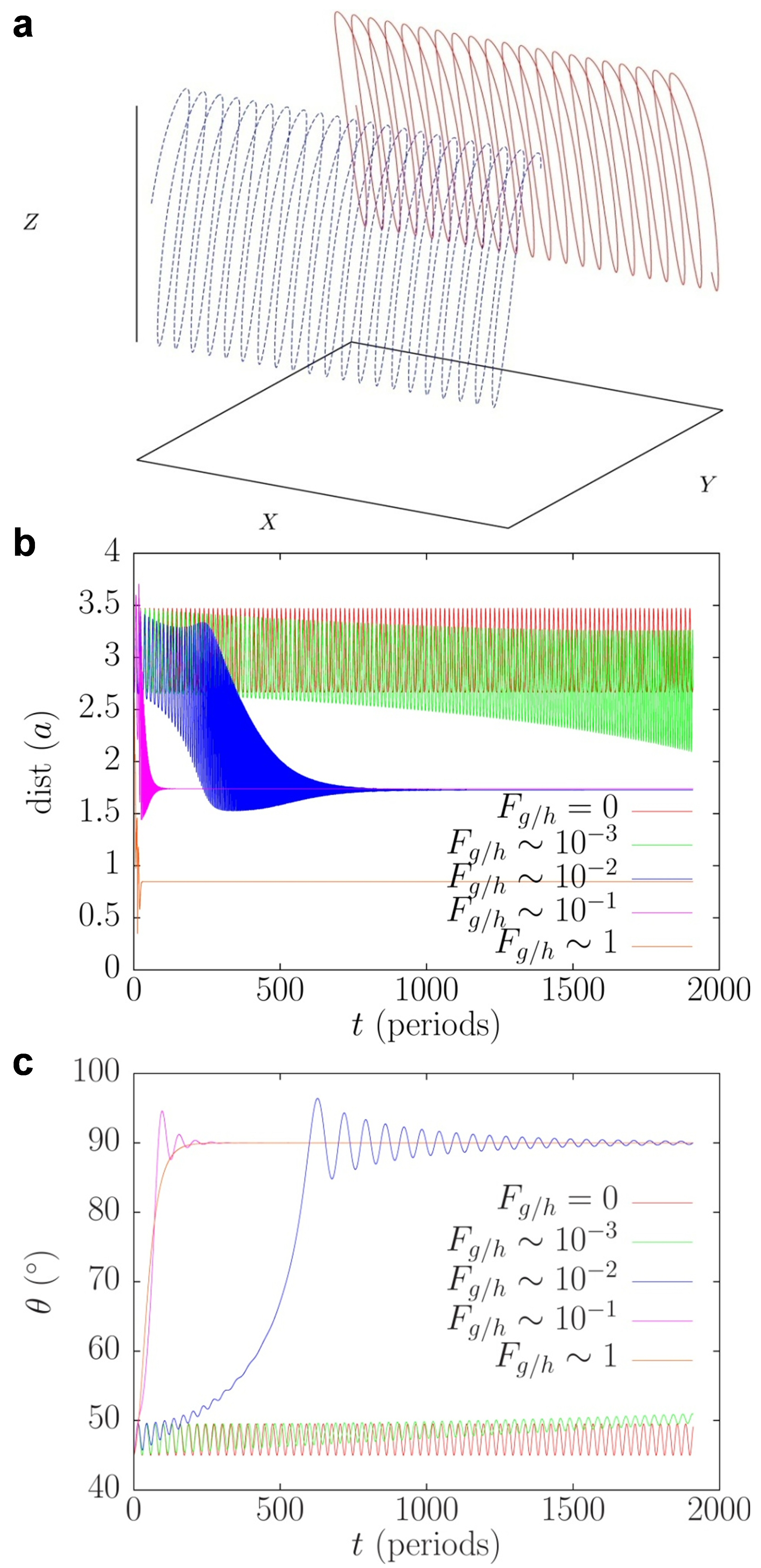}
\caption{\label{fig3} Hydrodynamic bound states. (\textbf{a}) 
Three dimensional trajectories of two particles (one in blue, the other in red) in a hydrodynamic bound state. The data are obtained from numerical simulations of the 
equations in the Supporting Information, 
with an initial angle $\vartheta = 45^o$ and at a distance of $d = 2.67a$, being $a$ the
particle radius, see MovieS4 in the Supplementary 
Information.
(\textbf{b,c}) Distance (b) and angle (c) 
between the two particles measured in term of the particle radius $a$ versus period of the driving
field for different values of $F_{g/h}$, 
the latter denotes the ratio between the gravitational $F_g$ and vicous $F_h$ forces.}
\end{center}
\end{figure*}

\begin{figure*}[t]
\begin{center}
\includegraphics[width=\textwidth]{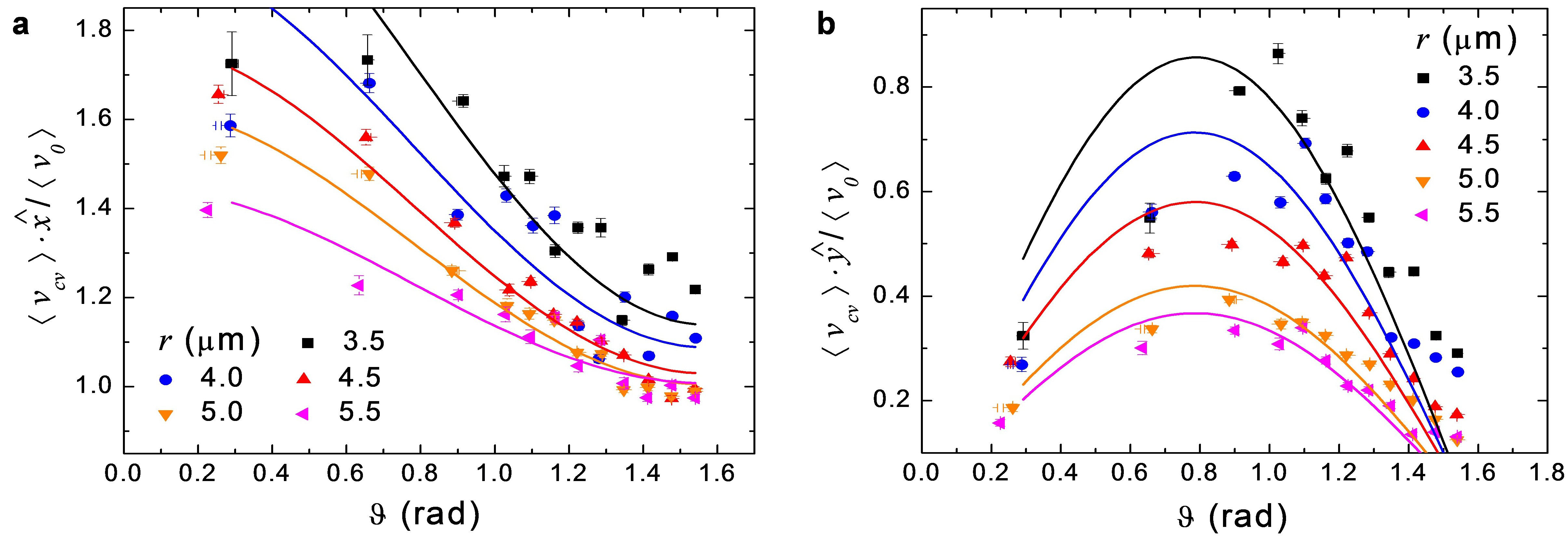}
\caption{\label{fig4} Average center of velocity of the pair. (\textbf{a},\textbf{b}) Components of 
of the velocity
$\langle v_{cv} \rangle$
along the $\hat{x}$ (a)
and $\hat{y}$ (b) directions,
versus angle $\vartheta$.
The pair of propellers forming the bound states 
is driven by
a rotating field with amplitude
$H_0=4400 {\rm A m^{-1}}$
and frequency $\omega= 502.6 {\rm rad  s^{-1}}$.
$\langle v_0 \rangle= 10.3 {\rm \mu m s^{-1}}$
denotes the speed of a single propeller driven by the same field,
$r$ is the center to center distance.
Scattered symbols denote experimental data,
continuous lines are fits following the
model developed in the text.}
\end{center}
\end{figure*}

\begin{figure*}[t]
\begin{center}
\includegraphics[width=\textwidth]{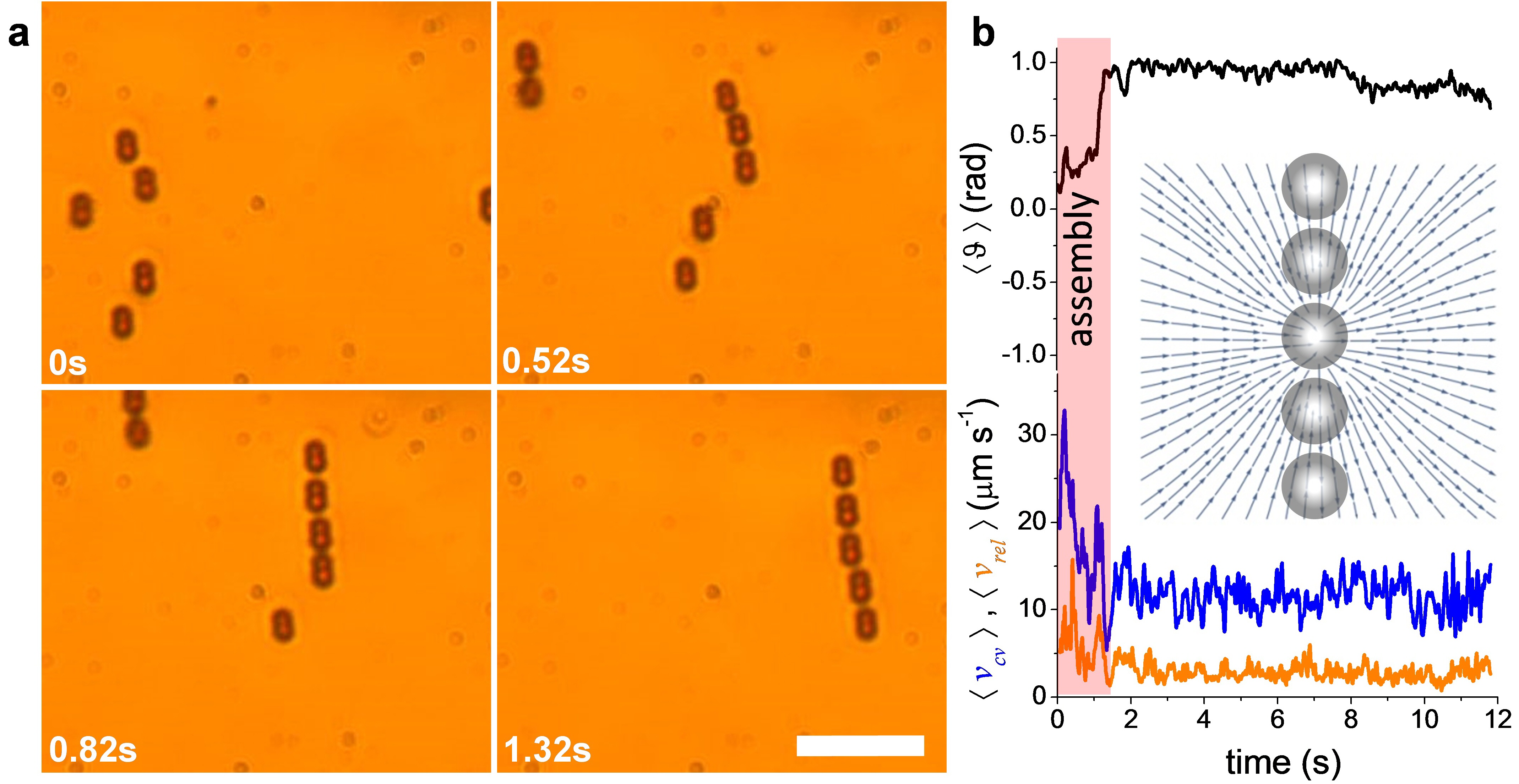}
\caption{\label{fig5} Self-assembly due to sole HI. (\textbf{a}) Sequence of images showing the formation of
a chain of $5$ propellers under a rotating field
with amplitude
$H_0=4400 {\rm A m^{-1}}$
and frequency $\omega= 502.6 {\rm rad \, s^{-1}}$.
Scale bar is $10 {\rm \mu m}$,
the corresponding video, MovieS3, is in the Supplementary 
Information.
(\textbf{b}) Evolution with time of the
average positional angle $\vartheta$ (top),
$\langle v_{cv}\rangle$ in  blue and $\langle v_{rel} \rangle$ in orange
(bottom).
The shaded red region in the graph denotes the assembly stage.
Inset illustrates the
flow velocity $u_x$
calculated at the same elevation of a chain
composed of $5$
propellers.}
\end{center}
\end{figure*}

\end{document}